\documentclass[conference]{IEEEtran}
\IEEEoverridecommandlockouts
\usepackage{cite}
\usepackage{url}
\usepackage{amsmath,amssymb,amsfonts}
\usepackage[table]{xcolor}
\usepackage{algorithm}
\usepackage{algorithmic}
\usepackage{hyperref}
\usepackage{censor}
\usepackage{booktabs}
\usepackage{graphicx,subfig}
\usepackage{enumerate}
\usepackage{textcomp}
\usepackage{todonotes}
\usepackage{xcolor}
\usepackage{amsthm}

\usepackage[normalem]{ulem}
\usepackage{listings}
\usepackage{colortbl}
\usepackage{tabularx}
\usepackage{caption}
\usepackage{enumitem}
\usepackage{multirow}
\usepackage{subcaption}
\usepackage{orcidlink}

\newif\ifanonymized
\anonymizedfalse

\def\BibTeX{{\rm B\kern-.05em{\sc i\kern-.025em b}\kern-.08em
    T\kern-.1667em\lower.7ex\hbox{E}\kern-.125emX}}
\begin{document}

\title{Pricing-Driven Resource Allocation in the Computing Continuum}

\ifanonymized
	\author{
		Anonymous Authors\\[0.2cm]

		\IEEEauthorblockA{
			Anonymous Affiliations\\
			Anonymous E-Mail
		}
	}
\else
	\author{
		\IEEEauthorblockN{
			Alejandro García-Fernández\IEEEauthorrefmark{1}\orcidlink{0009-0000-0353-8891},
			Boris Sedlak\IEEEauthorrefmark{2}\orcidlink{0009-0001-2365-8265},
			José Antonio Parejo\IEEEauthorrefmark{1}\orcidlink{0000-0002-4708-4606}\\
			Pantelis Frangoudis\IEEEauthorrefmark{3}\orcidlink{0000-0001-6901-7714},
			Antonio Ruiz-Cortés\IEEEauthorrefmark{1}\textsuperscript{(1)}\orcidlink{0000-0001-9827-1834},
			Schahram Dustdar\IEEEauthorrefmark{2}\IEEEauthorrefmark{4}\orcidlink{0000-0001-6872-8821}
		}

		\IEEEauthorblockA{\IEEEauthorrefmark{1}
			\textit{SCORE Lab, I3US Institute},
			Universidad de Sevilla, Sevilla, Spain\\
			\{agarcia29, japarejo, aruiz\}@us.es
		}

		\IEEEauthorblockA{\IEEEauthorrefmark{2}
			\textit{Distributed Intelligence and Systems-Engineering Lab},
			Universitat Pompeu Fabra, Barcelona, Spain\\
			\{boris.sedlak, schahram.dustdar\}@upf.edu
		}

		\IEEEauthorblockA{\IEEEauthorrefmark{3}
			\textit{Distributed Systems Group},
			TU Wien, Vienna, Austria\\
			pantelis.frangoudis@dsg.tuwien.ac.at
		}

		\IEEEauthorblockA{\IEEEauthorrefmark{4}
			\textit{ICREA}, Universitat Pompeu Fabra, Barcelona, Spain
		}
	}
\fi

\maketitle

\begin{abstract}
Deploying applications across the computing continuum requires selecting infrastructure nodes from geographically distributed and heterogeneous environments while satisfying constraints (e.g., performance, location). This decision problem is an important facet of resource allocation. As infrastructures grow in scale and heterogeneity, the resulting decision space becomes inherently combinatorial. 
Existing approaches typically formulate this problem as a constrained optimization task using ad-hoc representations of infrastructure topologies and demand, which hinders generalization across solutions. 
In contrast, Software as a Service ecosystems address a structurally similar configuration problem through pricings –structures whose plans and add-ons implicitly define the configuration space of possible subscriptions. 
Building on this observation, this work explores the potential of pricings as general-purpose representations of configuration spaces, positioning them as a promising alternative for addressing configuration problems, such as resource allocation, across the computing continuum. 
To this end, the paper presents the following contributions: i) a pricing-based formulation of the resource allocation problem in the computing continuum, enabling infrastructure configuration spaces to be represented using pricings; ii) a workflow that leverages PRIME, a pricing analysis engine, to explore these spaces and compute cost-optimal deployments satisfying functional and non-functional constraints; iii) generation processes for synthetic infrastructure topologies and workload demands; and iv) a dataset comprising 9,600 precomputed resource allocation scenarios to support benchmarking.

\end{abstract}


\section{Introduction}
\label{sec:intro}

The computing continuum has transformed resource allocation into a complex decision problem \cite{moreschini2022cloud}. Service providers must deploy applications across geographically distributed and heterogeneous infrastructures, spanning edge, fog, and cloud nodes, while satisfying constraints related to performance, cost, location, and resource availability \cite{soumplis2022resource}. Formally, the problem entails selecting a subset of available infrastructure nodes (a.k.a configuration) whose aggregated capacities satisfy such constraints. As infrastructures grow in scale and heterogeneity, this decision space becomes inherently combinatorial \cite{mihaiu2025resource}.

In practice, these decisions take place in multi-vendor ecosystems, where combining nodes from different providers often yields more flexible, cost-effective, and resilient configurations than single-vendor alternatives. However, this increases complexity, as each provider may support interoperability with certain counterparties yet exclude others. 

To cope with this complexity, most studies formulate the problem as a constrained optimization task in which the objective is to identify a configuration that satisfies functional constraints (e.g. storage) while optimizing non-functional attributes (e.g. throughput) \cite{mihaiu2025resource}. These approaches typically introduce ad-hoc formulations to describe infrastructure topology, resource capabilities, and deployment constraints, resulting in a proliferation of domain-specific representations that hinder generalization and comparison across solutions \cite{Vergara2023}. Moreover, despite their practical relevance, inter-provider relationships are seldom captured in such formulations.

Interestingly, a structurally similar problem has been studied in the context of Software as a Service (SaaS). In SaaS ecosystems, providers structure their offerings through pricings, a structure consisting of various plans and optional add-ons, which implicitly define a configuration space capturing all valid combinations of service capabilities and commercial options \cite{garcia_caise_2024}. Recent work has shown that these pricing structures can be interpreted as machine-oriented artifacts (a.k.a iPricings) whose configuration spaces can be systematically explored to identify valid or optimal offerings under a set of constraints \cite{garcia_tsc_2026}. In this setting, the exploration of pricing alternatives can likewise be formulated as a constraint satisfaction problem \cite{garcia_caise_2025}; however, rather than introducing a new modeling language, this approach leverages an existing service-level artifact –the pricing itself– to represent the configuration space.

This observation motivates the central idea of this work: pricings can be interpreted as a general-purpose representation of configuration spaces, applicable beyond their traditional role in SaaS systems. In particular, the resource allocation problem in the computing continuum can be understood as an instance of such configuration problems. Building on this insight, this work presents the following original contributions:

\begin{enumerate}
    \item A novel formulation of the resource allocation problem in the computing continuum based on pricings, enabling the infrastructure configuration space to be represented using these structures.
    
    \item A workflow that leverages PRIME \cite{garcia_caise_2025}, a pricing analysis engine, to explore the resulting pricing configuration space and compute cost-optimal deployments that satisfy both functional and non-functional requirements.
    
    \item As a part of our evaluation, we define processes to synthetically generate realistic infrastructure topologies and workload demands, enabling the construction of reproducible resource allocation scenarios.

    \item Using these processes, we provide a dataset containing 9,600 precomputed scenarios covering different topology sizes, demand levels, and optimization objectives, which can serve as a benchmark for future research \cite{LABPACK}.
\end{enumerate}

The remainder of this paper is structured as follows. Section \ref{sec:background} introduces foundational concepts of resource allocation and pricings. Section \ref{sec:pricing-driven-allocation} presents our proposed formulation and the associated deployment optimization workflow. Section \ref{sec:evaluation} reports on the evaluation and discusses potential threats to validity. Section \ref{sec:relatedWork} describes related work. Finally, Section \ref{sec:conclusions} draws conclusions and discusses future work.

\section{Background}
\label{sec:background}

To support the confluence of pricing mechanisms for optimizing the resource allocation, this section provides a comprehensive foundation for these two topics.

\subsection{Pricing}

\begin{figure}[htb]
    \captionsetup{justification=centering}
    \centering
    \includegraphics[width=\linewidth]{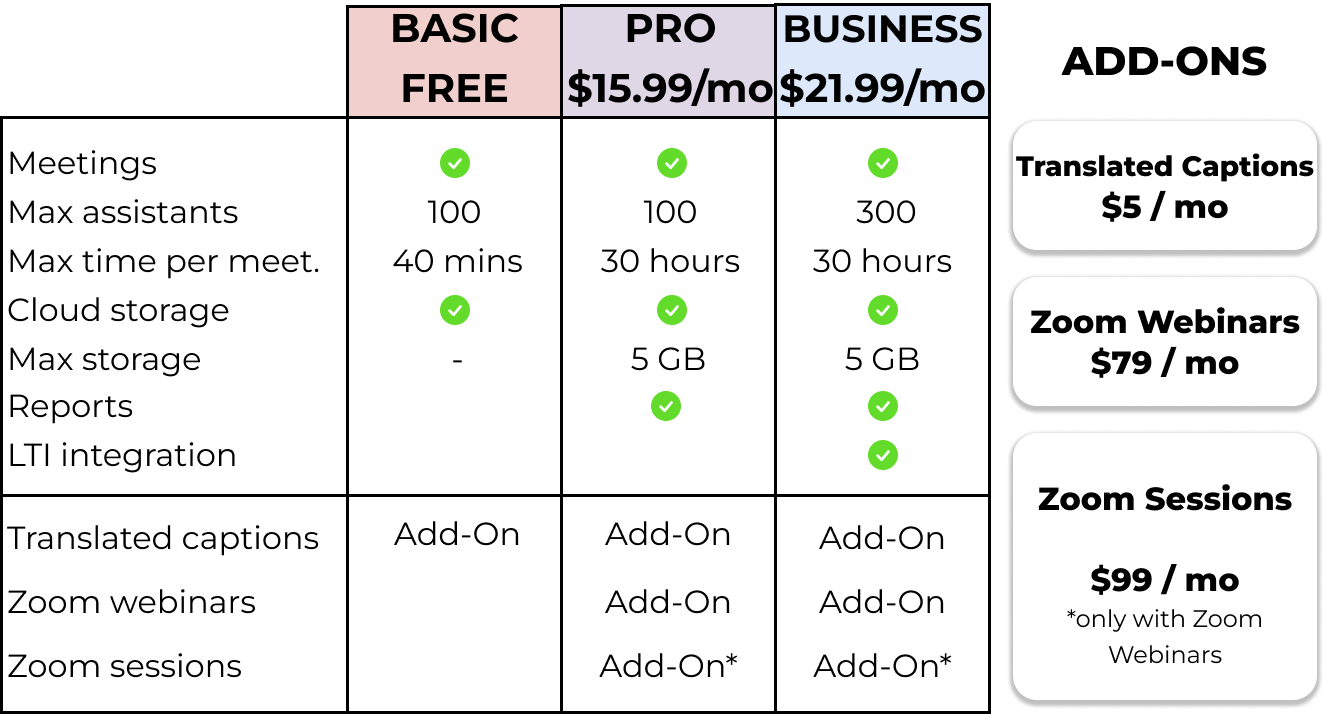}
    \caption{An example of a pricing: the Zoom platform. It contains 10 features, 3 usage limits, 3 plans, 3 add-ons}
    \label{fig:zoom-pricing}
\end{figure}

A pricing is a component of a SaaS customer agreement~\cite{molino2025integrating} that constrains user access to \textit{features} according to their subscription. A subscription typically consists of a \textit{plan} and an optional set of \textit{add-ons}, which determine the accessible features and may impose \textit{usage limits} based on measurable \textit{usage levels}~\cite{garcia_caise_2024}. The set of all valid subscriptions defines the software's \textit{configuration space}, making the pricing a compact representation of it~\cite{garcia_tsc_2026}. As an illustration, \figurename~\ref{fig:zoom-pricing} represents an excerpt of Zoom's pricing as of March 2026.

Recent work has extended this notion by elevating pricings from static business documents to first-class software artifacts, termed \emph{iPricings}~\cite{garcia_tsc_2026}. An iPricing provides a machine-oriented representation of a pricing, enabling –among other applications– automated analysis, validation, and optimization of configuration spaces.
To support these forms of automated reasoning, PRIME~\cite{garcia_caise_2025} was presented as one possible analysis engine. It maps iPricings to Constraint Satisfaction Optimization Problems (CSP), enabling their configuration spaces to be explored using CSP solvers such as Gecode~\cite{schulte2010modeling} or Choco~\cite{Prudhomme2022choco}. This has enabled, for example, (i) analyses of pricing evolution in SaaS ecosystems, revealing a growing gap between pricing variability and the capabilities of existing self-adaptation mechanisms~\cite{garcia_tsc_2026}, and (ii) the automatic validation of iPricings generated by automated modeling tools~\cite{cavero2026mint}.

\subsection{Resource Allocation}

Resource allocation in the computing continuum is a problem that spans several conceptual layers: discovery, filtering, optimization, and monitoring. Historically, most have been addressed in isolation. We briefly describe each below:

\begin{enumerate}[align=left, leftmargin=0pt, labelindent=\parindent,
listparindent=\parindent, labelwidth=0pt, itemindent=!]
    \item \textit{Discovery and Topology Construction.} At the lowest level, resource allocation requires discovering nodes (e.g., sensing or computing) and their capabilities. Due to its highly-distributed nature, Internet of Things (IoT) environments have a history of service discovery~\cite{7457097}. Common problems, as faced by \cite{9437637}, are the high degree of decentralization and the resource restrictions of IoT nodes. 

    \item \textit{Candidate Reduction and Heuristic Filtering.} To reduce the search space of potential services or nodes, candidates are first filtered \cite{zeng2004qos}. This intermediate reduction step –often referred to in the literature as resource selection or pre-selection– limits optimization to feasible candidates. In cloud–edge contexts, filtering commonly incorporates proximity, latency thresholds, or capability matching~\cite{mao2017survey}.

    \item \textit{Optimization and Allocation.} The core allocation problem is typically formulated as a constrained optimization problem. Early work in Web services considered multi-quality-of-service allocation under service-level agreement constraints \cite{xiong2006trust}. More recently, models for the cloud–edge continuum started to appear that allow a wide variety of objectives\cite{10646438}, such as optimizing energy consumption or overall cost.

    \item \textit{Monitoring and Adaptive Reallocation.} Once deployed, allocations must be continuously evaluated against service-level objectives. Edge computing literature emphasizes mobility-aware offloading and dynamic task migration under workload fluctuations \cite{mihaiu2025resource}.
\end{enumerate}

While these layers collectively describe the full allocation lifecycle, existing works model them independently. In particular, the optimization layer usually assumes that the configuration space is already given, without questioning how it is represented or structured. Our work departs from this assumption by explicitly modeling the configuration space itself as a first-class artifact.

\section{Pricing-Driven Resource Allocation}
\label{sec:pricing-driven-allocation}

This section formalizes the resource allocation problem in the computing continuum from a pricing-based perspective. The problem is defined in terms of offer\footnote{\textit{Offer} and \textit{topology} are used interchangeably, as the offer corresponds to the infrastructure topology.} ($\mathcal{O}$), demand ($\mathcal{D}$), and request ($\mathcal{R}$). This formulation is intentionally structured to support a systematic mapping onto the domain of pricings, enabling the exploration of the solution space and the derivation of cost-optimal allocations using pricing-driven techniques.

\subsection{Problem Definition}
\label{subsec:problem-definition}

We study a resource allocation scenario that consists of selecting a subset of infrastructure resources to satisfy a given workload at minimal cost, under technical, economic, and governance constraints. In this setting, we structure the problem around three complementary dimensions: (i) the offered infrastructure topology, which determines the space of potential resource combinations; (ii) the demand induced by users, which constrains this space based on required capacities; and (iii) a request, which captures additional constraints that are orthogonal to both offer and demand. While offer and demand jointly define the set of plausible allocations, the resulting search space grows exponentially with the number of available nodes, quickly becoming intractable at realistic scales. The request therefore plays a critical role by enabling further refinement of this space through explicit selection constraints—such as bounds on the number of nodes, budget limits, or provider-specific restrictions—thereby reducing the complexity of the optimization problem and guiding the search towards practically relevant solutions.

The following subsections formally define each of these dimensions and describe how they are modeled in our approach.

\subsubsection{Offer}

It is represented as an infrastructure topology, which captures both the available nodes on which software can be deployed and the constraints governing their joint use. Formally, we model the infrastructure topology as

\[
\mathcal{O} = (\mathcal{N}, \mathcal{B}),
\]

\noindent where $\mathcal{N}$ is a finite set of candidate nodes and $\mathcal{B}$  is a set of business rules, including interoperability constraints and mutual exclusions with other nodes or providers.
\noindent Each node $n \in \mathcal{N}$ is characterized by:

\begin{itemize}
    \item A \emph{resource capacity vector}
    \[
    \mathbf{r}(n) = \langle r_{\mathrm{ram}}, r_{\mathrm{sto}}, r_{\mathrm{cpu}}, r_{\mathrm{gpu}}, \ldots \rangle,
    \]
    which captures the set of resources exposed by the device, such as memory, network bandwidth, etc. The specific composition of this vector may be extended with any resource type for which capacity and cost information is available. We assume that such information has been obtained through prior discovery mechanisms.
    
    \item A \emph{context descriptor} for the node
    
    \[
    \mathbf{c}(n) = \langle \ell(n), v(n), p(n) \rangle,
    \]
    where $\ell(n)$ denotes the geographic location of the device, $v(n)$ its vendor, and $p(n)$ its rental price per time unit.
    
    \item A finite set of \emph{operational modes} $\mathcal{M}$, where each mode represents an alternative configuration of the same physical node with distinct resource limits and rental price.
\end{itemize}

\subsubsection{Demand}

captures the workload induced by end users as a consequence of interacting with a given application that runs over a subset of nodes in $O$. Formally, let $\mathcal{Z}$ denote a geographic zone of interest and let $U$ be the set of users located in such region. The interactions of users in $U$ with the target application induces a requirement for resources so as to deliver the service with a specified quality level. This demand is represented as an aggregate resource demand vector:
\[
\mathcal{D} = \langle \mathcal{D}_{\mathrm{ram}}, \mathcal{D}_{\mathrm{sto}}, \mathcal{D}_{\mathrm{cpu}}, \mathcal{D}_{\mathrm{gpu}}, \ldots \rangle,
\]
which specifies the total amount of resources required to sustain the expected workload generated by users in $\mathcal{Z}$.

A deployment configuration $C \subseteq \mathcal{N}$ is considered feasible with respect to demand if:
\[
\sum_{n \in C} \mathbf{r}(n) \succeq \mathcal{D},
\]
where $\succeq$ denotes component-wise dominance, that is, for every resource dimension \( i \),
\[
\sum_{n \in C} r_i(n) \ge \mathcal{D}_i.
\]

\noindent For instance, consider the following demand vector:

\[
\mathcal{D} = \langle \mathcal{D}_{ram}, \mathcal{D}_{sto}\rangle = \langle 8\mathrm{GB}, 32\mathrm{GB} \rangle
\]

\noindent and a configuration $C = \{n_1, n_2\}$, where 
\[
\mathbf{r}(n_1) = \langle 8\mathrm{GB}, 16\mathrm{GB} \rangle
\]
\[
\mathbf{r}(n_2) = \langle 2\mathrm{GB}, 64\mathrm{GB} \rangle
\]

The aggregated capacity of the configuration $C$ is then $\langle 10\mathrm{GB}, 80\mathrm{GB} \rangle$, which component-wise dominates $\mathcal{D}$. Therefore, $C$ satisfies the demand constraints.

\subsubsection{Request}

It represents additional constraints that may be defined to further restrict $\mathcal{O}$'s feasible configurations. We formalize these constraints as:
\[
\mathcal{R} = \langle N_{\max}, L_{\max}, P_{\max}, \mathcal{V} \rangle,
\]
where:
\begin{itemize}
    \item $N_{\max}$ is the maximum cardinality of a feasible deployment configuration, i.e. maximum number of nodes,
    \item $L_{\max}$ is the maximum admissible distance between a node and any point of the polygon $\mathcal{Z}$,
    \item $P_{\max}$ is the maximum admissible price per month,
    \item $\mathcal{V}$ is the set of allowed providers.
\end{itemize}

A configuration $C$ satisfies the request $\mathcal{R}$ if it respects the bounds on the number of selected nodes and the total price, and if all selected nodes belong to an allowed provider and lie within the admissible distance from the polygon $\mathcal{Z}$.
Formally:
\[
\begin{aligned}
\text{(C1)}\;& |C| \leq N_{\max}, \\
\text{(C2)}\;& \sum_{n \in C} p(n) \leq P_{\max}, \\
\text{(C3)}\;& \forall n \in C:\; v(n) \in \mathcal{V}, \\
\text{(C4)}\;& \forall n \in C:\; L(n,\mathcal{Z}) \leq L_{\max},
\end{aligned}
\]

\noindent where $L(n,\mathcal{Z})$ denotes the maximum distance between a node $n$ and any point of the polygon $\mathcal{Z}$. If $\ell(n)$ lies within $\mathcal{Z}$, the distance is defined as zero.

\subsubsection{Problem Statement}

Given an infrastructure topology $\mathcal{O}$, a demand vector $\mathcal{D}$, and a request $\mathcal{R}$, the resource allocation problem consists of finding a configuration $C^\ast \subseteq \mathcal{N}$ that satisfies demand and request constraints while minimizing the total deployment cost.

\subsection{Pricing-Driven Resource Allocation Workflow}
\label{subsec:approach}

To address the problem defined above, we propose a workflow that represents each topology configuration space as an iPricing and leverages PRIME (see Sec.~\ref{sec:background}) to analyze the resulting model in order to compute cost optimal deployment configurations that satisfy the specified \textit{demand} and \textit{request}.

\begin{figure*}[htb]
\centering
\includegraphics[width=\textwidth]{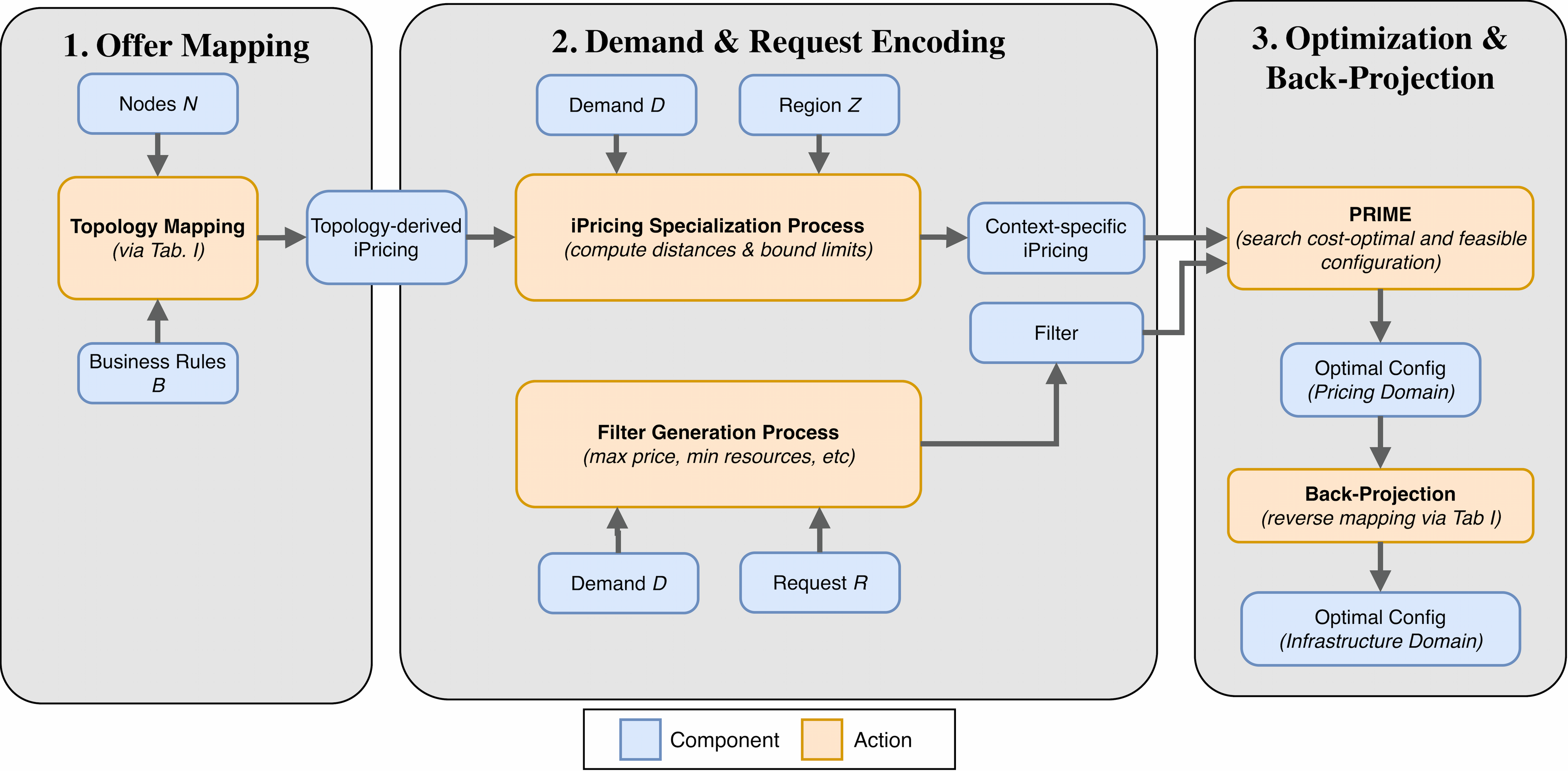}
\caption{Pricing-driven resource allocation workflow}
\label{fig:workflow}
\end{figure*}

\figurename~\ref{fig:workflow} summarizes the proposed workflow in three stages: (1) offer mapping, where $\mathcal{O}$ is projected onto the pricing domain; (2) demand and request encoding, where the pricing is specialized to a concrete problem instance and the search space is constrained; and (3) optimization and back-projection, where PRIME identifies a globally optimal configuration that is then mapped back to the infrastructure domain.

\subsubsection{Topology Mapping}

In this stage, the topology $\mathcal{O}$ is projected onto the pricing domain following the mapping defined in Tab.~\ref{tab:domain-mapping}. This projection yields an iPricing that compactly represents $\mathcal{O}$'s configuration space.

\begin{table}[htb]
\captionsetup{justification=centering}
\centering
\caption{Mapping between infrastructure and pricing domain}
\label{tab:domain-mapping}

\setlength{\tabcolsep}{10pt}      
\renewcommand{\arraystretch}{1.4} 

\begingroup
\rowcolors{2}{gray!10}{white}

\begin{tabularx}{\linewidth}{|>{\raggedright\arraybackslash}X|>{\raggedright\arraybackslash}X|}
\hline
\rowcolor{gray!25}
\multicolumn{1}{|c|}{\textbf{Infrastructure Domain}} &
\multicolumn{1}{c|}{\textbf{Pricing Domain}} \\
\hline
Currency & Currency \\
\hline
Node & Add-on \\
\hline
Node Types & Features \\
\hline
Topology Available Resources & Usage limits \\
\hline
Distance & Usage limit \\
\hline
Node Resources & Usage Limit Extension \\
\hline
Provider Exclusions & Add-on Exclusions \\
\hline
Resources Unitary Price & Price Expression \\
\hline
\end{tabularx}

\endgroup
\end{table}

\begin{figure}[htb]
    \centering
    \includegraphics[width=\linewidth]{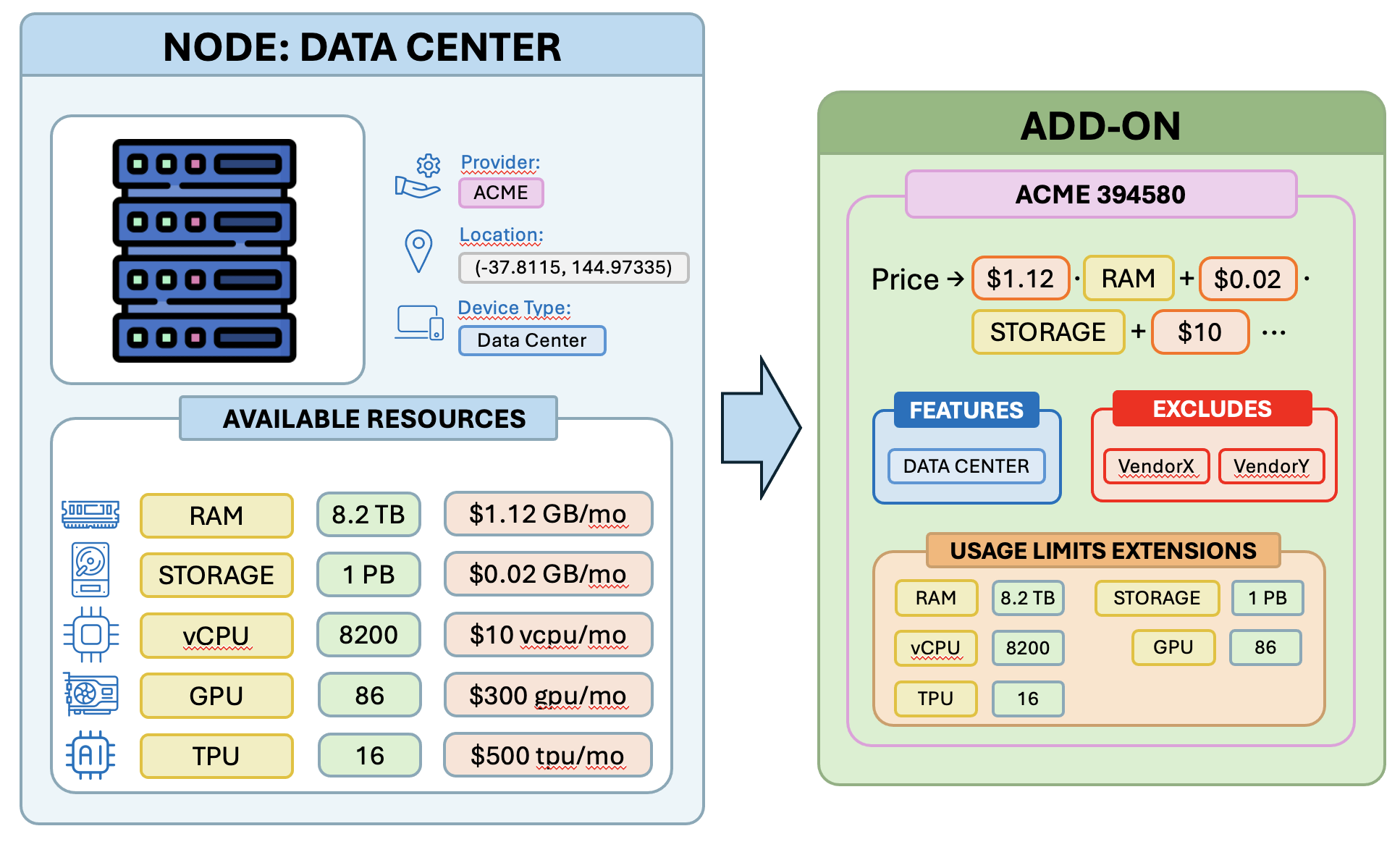}
    \caption{Mapping of a sample node into the pricing domain}
    \label{fig:data-center-mapping}
\end{figure}

However, the resulting iPricing cannot yet be directly processed by PRIME, as the monthly price of each add-on is not fully instantiated. Nodes in $\mathcal{N}$ typically expose more capacity than what is required by a concrete demand scenario; therefore, associating a fixed price to selecting a node would be misleading. Instead, unitary resource costs are considered to construct symbolic price expressions that define the cost of an add-on as a function of the effective resources it contributes. For example, a node offering memory at \$1.1/GB and storage at \$0.02/GB is associated with a price expression of the form:
\[
\textit{requested\_ram} \cdot 1.1 + \textit{requested\_storage} \cdot 0.02
\]
These expressions are left unresolved and will be instantiated in the next stage once demand information becomes available.
\figurename~\ref{fig:data-center-mapping} illustrates a sample outcome of this process.

\subsubsection{Demand and Request Encoding}

In this stage, the demand vector $\mathcal{D}$, the region $\mathcal{Z}$, and the request $\mathcal{R}$ are jointly used to (i) specialize the topology-derived iPricing to the concrete deployment context, and (ii) derive a filter that constrains the configuration space explored by PRIME.

Regarding (i), transformations are driven by $\mathcal{D}$ and $\mathcal{Z}$. For each infrastructure node $n \in N$, the distance $L(n, \mathcal{Z})$ is computed and incorporated into the pricing as a usage limit. In addition, usage limit extensions associated with each add-on are bounded by the corresponding demand requirements. Formally, for each add-on $a$ and each resource dimension $i$, the effective usage limit extension is defined as
\[
u_i^{a} = \min \left( u_{i,\mathrm{0}}^{a}, \, \mathcal{D}_i \right),
\]
ensuring that no add-on contributes capacity beyond what is required by the demand instance. Based on these bounded usage limits, the price expressions of each add-on are then evaluated, yielding a concrete monthly price that reflects the effective resources contributed by selecting that node. For instance, if we consider the following demand vector: 

\[
\mathcal{D} = \langle \mathcal{D}_{ram}, \mathcal{D}_{sto} \rangle = \langle 20, 100 \rangle
\]

\noindent the price expression from (1) would be instantiated as follows:

\[
20 \cdot 1.1 + 100 \cdot 0.02 = \$24
\]

Regarding (ii), the filter is obtained by combining $\mathcal{D}$ and $\mathcal{R}$, with the former inducing minimum usage limits and the latter bounding the total price and the configuration cardinality.

\subsubsection{Optimization and Back-Projection}

Artifacts generated in (2) are provided to PRIME \cite{garcia_caise_2025}, which first uses them to delimit the admissible configuration space, and then searches this space for an optimal configuration that satisfies all requirements while minimizing the total monthly cost.

Finally, the resulting solution is mapped back to the infrastructure domain by applying Tab.~\ref{tab:domain-mapping} in reverse, yielding the optimal set of devices on which to deploy the infrastructure.

\section{Evaluation}
\label{sec:evaluation}

The goal of our evaluation is to assess both the modeling capabilities and the computational viability of our proposal. To this end, we defined the following research questions (RQs):

\begin{itemize}
    \item \textit{\textbf{RQ$_1$ (Modeling Expressiveness)}: Is the proposed pricing-based formulation expressive enough to model realistic multi-provider scenarios involving interoperability constraints and heterogeneous infrastructure?}
    \item \textit{\textbf{RQ$_2$ (Optimization Capabilities)}: Can the approach identify cost-optimal deployments that satisfy a set of given constraints within reasonable time constraints?}
    \item \textit{\textbf{RQ$_3$ (Scalability)}: How does the performance of the approach scale as the complexity of the deployment scenario increases gradually?}
\end{itemize}

\subsection{Experimental Setup}
\label{sec:experimental-setup}

\subsubsection{\textbf{Dataset}}
\label{sec:dataset}

Our evaluation is based on the Edge User Allocation (EUA) dataset~\cite{lai_phulaieua-dataset_2025}, which provides the geographic distribution of 95{,}562 edge servers (hereinafter, nodes) across Australia –primarily in the Melbourne metropolitan area– together with the location of 4{,}748 clients. Since our goal is to generate realistic topologies rather than to model user behavior directly, we rely exclusively on the infrastructure-related information in the dataset. The rationale behind discarding client-related data is discussed in more detail in Sec.~\ref{sec:demand-generation}.

Given this objective, most of the original node attributes are discarded, retaining only: \textit{latitude}, \textit{longitude}, \textit{elevation}, and the node \textit{name}. The resulting dataset is then preprocessed to incorporate some information required for the evaluation:

\begin{itemize}[align=left, leftmargin=0pt, labelindent=\parindent,
listparindent=\parindent, labelwidth=0pt, itemindent=!]
    \item \textit{Provider inference.} The \textit{name} attribute is used to identify nodes containing information about their provider, restricted to: Optus, Telstra, Vodafone, Macquarie, and Telecom. Out of the 95{,}562 nodes in the dataset, 18{,}822 contain such information, which we consider sufficient for evaluation purposes. Once the inference is completed, the original \textit{name} attribute is discarded and replaced by an explicit \textit{provider} field.
    
    \item \textit{Synthetic resource generation.} As the EUA dataset does not include hardware specifications, each node is further enriched with a synthetic resource capacity vector ($\mathbf{r}(n)$). To this end, nodes are randomly assigned to one of three infrastructure tiers –edge, fog, or cloud– reflecting increasing levels of computational capacity. For each tier, resource capacities are generated pseudo-randomly using a predefined configuration to ensure reproducibility.\footnote{Due to space constraints, further details of the generation procedure are omitted and made available in the accompanying artifact~\cite{LABPACK}.} We consider the following resource dimensions: RAM, storage, CPU, GPU, and TPU. Resource prices are assigned per unit and may vary across providers and infrastructure tiers.
\end{itemize}

\subsubsection{\textbf{Synthetic Topology Generation}}
\label{sec:topology-generation}

Synthetic topologies are generated by spatially sampling the enriched EUA dataset (see Sec.~\ref{sec:dataset}). Each topology is defined by specifying the geographic coordinates (latitude, longitude, and elevation) of a center point within the Melbourne area, together with a radius that determines a three-dimensional spherical region. All nodes whose location falls within this region are selected as candidates for the topology.

In addition to the spatial parameters, a maximum number of nodes may be provided to cap the size of the resulting topology by randomly selecting a subset of nodes. In parallel, a set of business rules is specified to capture interoperability constraints and exclusion agreements among providers.

\begin{figure}[htb]
    \centering
    \includegraphics[width=\linewidth]{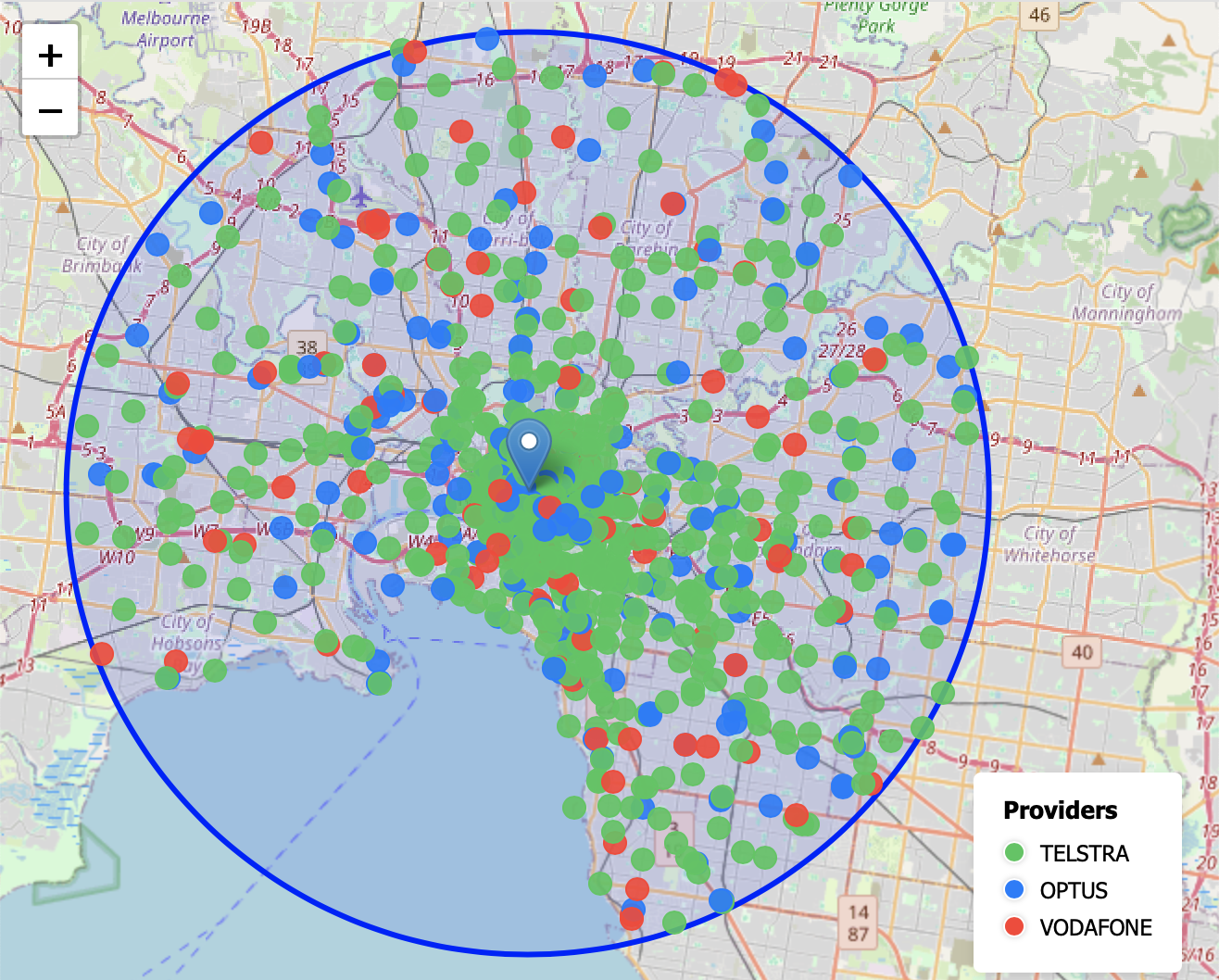}
    \caption{Sample topology generated around the Melbourne city center. Each point represents a node. Figure adapted from \cite{LABPACK}}
    \label{fig:sample-topology}
\end{figure}

Finally, the selected nodes and the associated business rules are used as input to the ``topology mapping'' stage of our approach (see Sec~\ref{subsec:approach}), yielding an iPricing that represents the generated topology in the pricing domain. \figurename~\ref{fig:sample-topology} illustrates a sample topology produced using this process.

\subsubsection{\textbf{Synthetic Demand Generation}}
\label{sec:demand-generation}

As discussed in Sec~\ref{sec:dataset}, we do not use the client-related information provided by the EUA dataset to derive demand. The dataset only reports the geographic location of clients, without any indication of the workload they generate or the quality-of-service requirements they impose. Moreover, even if we consider all clients in the dataset (4{,}748), this number is insufficient to represent large-scale deployments.

Instead, demand is generated synthetically in accordance with our problem formulation, which represents it as an aggregate resource demand vector $\mathcal{D}$ associated with a region of interest $\mathcal{Z}$. To this end, we define a function $f_{\mathrm{res}}$ that maps the number of users $N = |U|$ assumed to be located within $\mathcal{Z}$ to a demand vector:
\[
\mathcal{D} = f_{\mathrm{res}}(N) =
\langle
\mathcal{D}_{\mathrm{ram}},
\mathcal{D}_{\mathrm{sto}},
\mathcal{D}_{\mathrm{cpu}},
\mathcal{D}_{\mathrm{gpu}},
\mathcal{D}_{\mathrm{tpu}}
\rangle
\]

The objective of $f_{\mathrm{res}}$ is not to predict the exact requirements of a production deployment, but to generate \emph{plausible, internally consistent, and reproducible} demand instances for evaluating our proposal across different scenarios.

\begin{itemize}[align=left, leftmargin=0pt, labelindent=\parindent,
listparindent=\parindent, labelwidth=0pt, itemindent=!]

    \item \textit{User Activity and Throughput Characterization:} Not all users are active simultaneously. Let $N$ denote the total number of users and $N_a$ the number of active users. We model $N_a$ as a binomial random variable:
    \[
    N_a \sim \mathrm{Binomial}(N, p_s)
    \]
    where $p_s \in [0,1]$ denotes the service concurrency level, a standard abstraction in capacity planning~\cite{menasce2002capacity}.
    
    Each active user generates requests at a certain rate. The per-user request rate is modeled as a random variable with a configurable mean, allowing moderate variability across users. The aggregate request arrival rate $\lambda$ (requests per second) is then computed as:
    \[
    \lambda = \alpha_s \sum_{i=1}^{N_a} r_i
    \]
    where $r_i$ denotes the request rate of user $i$ and $\alpha_s \geq 1$ is a service-specific safety factor accounting for burstiness and modeling uncertainty, following established practices in large-scale system provisioning~\cite{barroso2013datacenter}.

    \item \textit{Memory Demand:} is decomposed into three service-specific components: a fixed baseline footprint, a shared component amortized across users, and a per-session overhead:
    \[
    \mathcal{D}_{\mathrm{ram}} =
    \alpha_s \left(
    \mathcal{D}^{\mathrm{base}}_{\mathrm{ram},s}
    + \mathcal{D}^{\mathrm{shared}}_{\mathrm{ram},s}(N_a)
    + \mathcal{D}^{\mathrm{session}}_{\mathrm{ram},s} \cdot N_a
    \right)
    \]
    
    Memory usage is not assumed to scale linearly with the number of users. Empirical studies show that large portions of memory consumption –such as caches, shared models, or runtime state– are amortized across sessions~\cite{shen2011cloudscale}. Accordingly, the shared component is modeled as:
    \[
    \mathcal{D}^{\mathrm{shared}}_{\mathrm{ram},s}(N_a) = \beta_s \cdot \sqrt{N_a}
    \]
    where $\beta_s$ is a service-specific scaling coefficient.

    \item \textit{Compute and Accelerator Demand:} Compute resources are dimensioned based on throughput rather than per-user provisioning; accordingly, CPU demand is modeled as follows:
    \[
    \mathcal{D}_{\mathrm{cpu}} =
    \frac{\lambda \cdot t^{\mathrm{cpu}}_s}{u^{\mathrm{cpu}}_s}
    \]
    where $t^{\mathrm{cpu}}_s$ is the average CPU service time per request and $u^{\mathrm{cpu}}_s \in (0,1)$ is the target utilization threshold.
    
    Accelerator demand is computed analogously. Let $\phi^{\mathrm{gpu}}_s$ denote the fraction of requests executed on the GPU:
    \[
    \mathcal{D}_{\mathrm{gpu}} =
    \phi^{\mathrm{gpu}}_s \cdot \frac{\lambda \cdot t^{\mathrm{gpu}}_s}{u^{\mathrm{gpu}}_s}
    \]
    
    TPU demand is computed using the same formulation, with service-specific parameters adjusted to TPU execution.~In all cases, accelerator demand is expressed in \emph{equivalent processing units}, capturing sustained throughput rather than physical devices, and reflecting batching and multiplexing effects commonly observed in modern inference-serving systems~\cite{crankshaw2017clipper}.

    \item \textit{Storage Demand:} as with memory demand, it is decomposed into three components: a fixed baseline, a per-session overhead, and an optional log retention term:
    \[
    \mathcal{D}_{\mathrm{sto}} =
    \alpha_s \left(
    \mathcal{D}^{\mathrm{base}}_{\mathrm{sto},s}
    + \mathcal{D}^{\mathrm{session}}_{\mathrm{sto},s} \cdot N_a
    + \lambda \cdot L_s \cdot W_s
    \right)
    \]
    
    In this formulation, $L_s$ denotes the service-specific log volume generated per request and $W_s$ the retention window.
\end{itemize}

Once demand has been generated independently for each resource dimension, the resulting values are combined to construct the aggregate demand vector $\mathcal{D}$. Further details on the specific values assigned to parameters in our evaluation are provided in the accompanying artifact~\cite{LABPACK}.

\subsection{Experimental Design}
\label{sec:experimental-design}

Our evaluation is based on a single, systematic experiment that uses the topology and demand generation strategies introduced in Sec.~\ref{sec:experimental-setup} to construct a set of realistic deployment scenarios that vary along three orthogonal dimensions: the type of the application to be deployed, the size of the offered topology (in terms of candidate nodes), and the intensity of the demand (in terms of the number of clients). Their combination yields deployment scenarios of increasing complexity on which the proposed pricing-driven workflow is evaluated. In total, the experiment comprises 9{,}600 distinct test cases.

\paragraph{Relation to research questions}
RQ1 is addressed by assessing whether all deployment instances defined in the experiment can be represented using the proposed pricing-based formulation. RQ2 is evaluated by analyzing the proportion of test cases for which the approach successfully identifies a feasible and cost-optimal configuration. Finally, RQ3 is addressed by measuring the execution time required to solve each instance and examining how it evolves as the number of clients or candidate infrastructure nodes increases.

\paragraph{Application scenarios}
We consider four application scenarios, each deployed over a different area of the Melbourne region and characterized by distinct requirements:

\begin{itemize}
    \item \textbf{A CCTV surveillance system} deployed around the Royal Melbourne Hospital. This scenario is characterized by a sustained high data throughput due to continuous video streams, combined with moderate latency requirements.
    
    \item \textbf{Virtual Reality (VR) service} deployed in the Melbourne city center to support interactive touristic experiences. This scenario requires ultra-low latency and high data rates to ensure an acceptable quality of experience.
    
    \item \textbf{Robotic arm coordination system} deployed in the Port of Melbourne to coordinate multiple robotic arms used for logistics and cargo handling. This scenario features strict latency constraints but comparatively lower data volumes.
    
    \item \textbf{LiDAR-based sensing and data collection system} deployed in the Sunbury area to support agricultural monitoring tasks. This scenario is dominated by very large data volumes, with less stringent latency requirements.
\end{itemize}

This setting follows the multi-scenario deployment perspective adopted by \cite{jin_ms-gd-p_2024}, which evaluated placement strategies under heterogeneous demand and service requirements.

\paragraph{Deployment scales and scalability dimensions}
Beyond the application dimension, scalability in our experiment is explored by independently varying the size of the offered topology and the intensity of the demand. To this end, we define three scales: \emph{small} (S), \emph{medium} (M), and \emph{large}~(L).

For each application and deployment scale, we instantiate four increasing demand levels (number of clients) and four increasing bounds on the number of candidate nodes. Their values are derived from the ranges reported in Tab.~\ref{tab:experimental-scenarios}, yielding a total of 96 distinct scenario types. For each scenario type, we then generate 100 topologies (see Sec.~\ref{sec:topology-generation}) together with corresponding demand workloads that satisfy the imposed constraints. This repeated sampling strategy enables a robust estimation of execution time by averaging results across structurally different yet semantically equivalent problem instances.

\begin{table}[htb]
\centering
\caption{Experimental scenarios considered in the evaluation. For each application and deployment scale, ranges indicate the minimum and maximum values used to generate four uniformly spaced configurations for the number of clients and candidate nodes. Node ranges depend only on the deployment scale, while user ranges are application-specific.}
\label{tab:experimental-scenarios}

\setlength{\tabcolsep}{8pt}
\renewcommand{\arraystretch}{1.2}

\begin{tabular}{|c|c|c|c|}
\hline
\textbf{Scale} & \textbf{Application} & \textbf{Max. Users} & \textbf{Max. Nodes} \\
\hline
\multirow{4}{*}{S}
  & CCTV   & 20--80   & \multirow{4}{*}{5--30} \\ 
  & VR     & 25--100  &  \\ 
  & Robot  & 15--60   &  \\ 
  & LiDAR  & 50--200  &  \\
\hline
\multirow{4}{*}{M}
  & CCTV   & 100--400  & \multirow{4}{*}{50--200} \\ 
  & VR     & 200--800  &  \\ 
  & Robot  & 75--300   &  \\ 
  & LiDAR  & 500--2000 &  \\
\hline
\multirow{4}{*}{L}
  & CCTV   & 500--2000  & \multirow{4}{*}{300--500} \\ 
  & VR     & 1500--8000 &  \\
  & Robot  & 250--1000  &  \\
  & LiDAR  & 2500--5000 &  \\
\hline
\end{tabular}
\end{table}

To isolate the impact of each scalability dimension when addressing RQ3, the two variables are varied independently. When analyzing scalability with respect to demand, the maximum number of candidate nodes is kept fixed to 20. Conversely, when analyzing scalability with respect to topology size, the number of clients is fixed to 100.

\begin{table*}[htb]
\centering
\caption{Request constraints applied per application scenario and deployment scale. Distance bounds limit the maximum admissible distance from nodes to the region of interest $\mathcal{Z}$. Max.\ nodes bound the size of the resulting deployment configuration.}
\label{tab:request-constraints}

\setlength{\tabcolsep}{6pt}
\renewcommand{\arraystretch}{1.15}

\begin{tabular}{|c|c|c|c|c|}
\hline
\textbf{Scale} & \textbf{Application} & \textbf{Max. Distance (m)} & \textbf{Max. Nodes} & \textbf{Allowed Node Types} \\
\hline
\multirow{4}{*}{S}
 & VR      & 3{,}750   & 4 & CAMERA, SENSOR, NETWORK\_NODE, DATA\_CENTER \\
 & Robot   & 500       & 3 & SENSOR, COMPUTER, NETWORK\_NODE \\
 & LiDAR   & 2{,}500   & 3 & SENSOR, NETWORK\_NODE, DATA\_CENTER \\
 & CCTV    & 500{,}000 & 3 & CAMERA, NETWORK\_NODE, DATA\_CENTER \\
\hline
\multirow{4}{*}{M}
 & VR      & 7{,}500   & 6 & CAMERA, SENSOR, NETWORK\_NODE, DATA\_CENTER \\
 & Robot   & 1{,}000   & 6 & SENSOR, COMPUTER, NETWORK\_NODE \\
 & LiDAR   & 5{,}000   & 6 & SENSOR, NETWORK\_NODE, DATA\_CENTER \\
 & CCTV    & 800{,}000 & 6 & CAMERA, NETWORK\_NODE, DATA\_CENTER \\
\hline
\multirow{4}{*}{L}
 & VR      & 15{,}000  & 10 & CAMERA, SENSOR, NETWORK\_NODE, DATA\_CENTER \\
 & Robot   & 1{,}500   & 10 & SENSOR, COMPUTER, NETWORK\_NODE \\
 & LiDAR   & 10{,}000  & 10 & SENSOR, NETWORK\_NODE, DATA\_CENTER \\
 & CCTV    & 1{,}000{,}000 & 10 & CAMERA, NETWORK\_NODE, DATA\_CENTER \\
\hline
\end{tabular}
\end{table*}

\paragraph{Provider configuration and additional constraints}
All experiments consider three infrastructure providers: TELSTRA, OPTUS, and VODAFONE. Interoperability constraints are enforced such that VODAFONE nodes can interoperate with both TELSTRA and OPTUS, while TELSTRA and OPTUS are mutually exclusive. It is important to note that they are fictitious and defined solely to create representative evaluation scenarios for the case study.

In addition, each test case is further restricted by a request that captures deployment-specific requirements. These include: (i) a maximum admissible distance between selected nodes and $\mathcal{Z}$, used as a proxy for latency constraints; (ii) an upper bound on the number of nodes that may compose the final deployment configuration; (iii) a maximum budget; and (iv) constraints on the types of nodes that may be selected, reflecting application-specific functional requirements.

These request parameters vary across application scenarios and deployment scales and are summarized in Tab.~\ref{tab:request-constraints}.

\subsection{Discussion of Results}
\label{sec:discussion}

All experiments were executed on a Mac Mini machine equipped with an Apple M4 Pro processor and 24 GB of RAM. The following discussion analyzes the obtained results in relation to the defined research questions.

\paragraph{$RQ_1$. Modeling Expressiveness}
The results provide strong evidence that the proposed pricing-based formulation is sufficiently expressive to model realistic multi-provider resource allocation scenarios. All 9{,}600 problem instances presented in Sec.~\ref{sec:experimental-design} were successfully represented as iPricings without requiring ad-hoc adaptations. This was achieved through a programmatic generation of pricing configurations using the python reference implementation available at \cite{LABPACK}.

\paragraph{$RQ_2$. Optimization Capabilities}
The reference implementation was able to compute a feasible and cost-optimal deployment for \emph{all} test cases considered in the experiment.

In addition, we explicitly evaluated infeasible configurations by constructing request profiles whose constraints could not be jointly satisfied by the available infrastructure. In such cases, the solver consistently reported the absence of a solution, indicating that the approach correctly distinguishes between feasible and infeasible instances rather than failing silently or returning suboptimal configurations. 

\begin{figure*}[htb]
    \centering
    \subfloat[Median execution time per \#clients scale]{
        \includegraphics[width=0.42\textwidth]{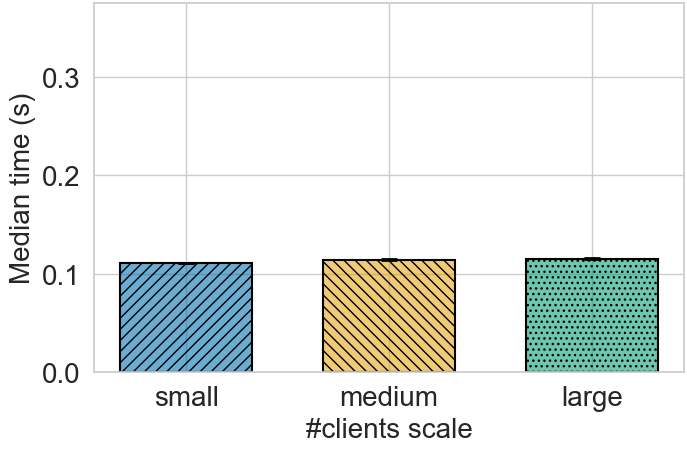}
        \label{fig:results-clients}
    }
    \hfill
    \subfloat[Median execution time vs increasing number of candidate nodes]{
        \includegraphics[width=0.51\textwidth]{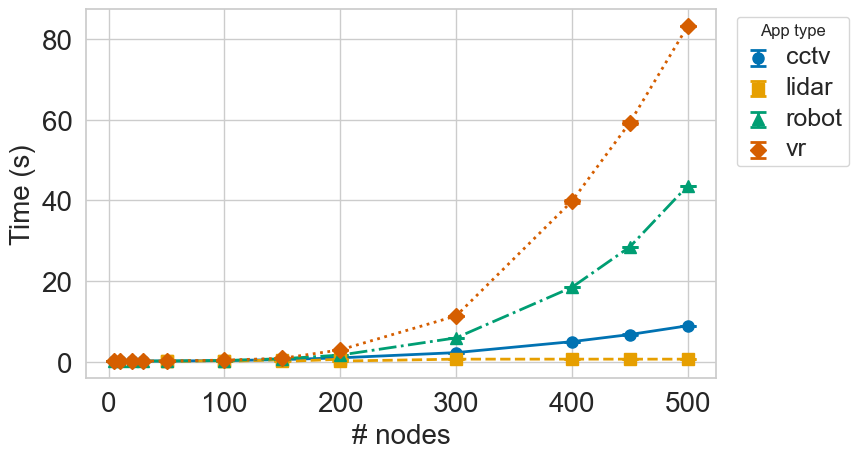}
        \label{fig:results-nodes}
    }
    \caption{Execution time of the proposed approach under increasing problem size. Each point represents the median solving time computed over 100 independent topology instances generated for the corresponding experimental configuration. Error bars denote $95\%$ confidence intervals around the median. Due to the low intra-scenario variability observed in the experiments, these intervals are often visually negligible. A detailed statistical analysis is provided in \cite{LABPACK}.}
    \label{fig:results}
\end{figure*}

Execution time of the proposed approach under increasing problem size. Each point represents the median solving time computed over 100 independent topology instances generated for the corresponding experimental configuration. Error bars denote $95\%$ confidence intervals around the median. Due to the low intra-scenario variability observed in the experiments, these intervals are often visually negligible. A detailed statistical analysis of runtime stability and distributional properties is provided in the accompanying lab package.

\paragraph{$RQ_3$. Scalability}
\figurename~\ref{fig:results} illustrates the evolution of execution time as the two scalability dimensions considered –i.e the number of clients and candidate nodes– increase. Each value correspond to the median execution time obtained across the 100 topology instances generated for each scenario.

On the one hand, \figurename~\ref{fig:results-clients} demonstrates that execution time remains stable as the number of clients grows. This behavior is expected, since scaling demand only tightens constraints (e.g., higher usage limits), but does not introduce additional decision variables nor expand the search space. In short, demand impact on execution time is negligible.

On the other hand, \figurename~\ref{fig:results-nodes} shows that execution time increases with the number of candidate nodes, as expected due to the expansion of the decision space. However, this growth is scenario-dependent. Applications with more specialized resource requirements –such as VR, which relies on relatively scarce resources like GPUs– exhibit steeper increases than scenarios dominated by widely available resources (e.g., storage). This indicates that solver performance is strongly influenced by the availability and distribution of the required resources.

Despite this trend, execution times remain within practical bounds. Problem instances involving up to 200 candidate nodes are consistently solved in \textit{under three seconds}, making the approach suitable for real-world planning and decision-support settings. Even for larger infrastructures, execution times stay in the order of seconds rather than minutes.


Overall, these results show that the proposed approach is expressive enough to model all the considered multi-provider scenarios ($RQ_1$), consistently identifies feasible and cost-optimal solutions when they exist ($RQ_2$), and scales favorably to realistic infrastructure sizes, handling a broad range of deployment scenarios within practical time limits ($RQ_3$).

\subsection{Threats to Validity}
Despite the positive results obtained in our evaluation, several threats to validity must be acknowledged.

\textbf{Internal Validity.} The main threat to internal validity relates to the synthetic generation of topologies and demand workloads. 
While these values were designed to be plausible and were grounded in established capacity planning abstractions, they may not perfectly capture the complex, non-linear resource dynamics of production-grade scenarios.

\textbf{External Validity.} The generalizability of our findings is limited by the use of a single geographically concentrated dataset (EUA) focused on the Melbourne metropolitan area. Although we evaluated 9{,}600 test cases across 96 distinct scenario types, different infrastructure distributions or provider ecosystems with more complex interoperability rules might yield different performance results. Furthermore, our evaluation considered a fixed set of three providers, which may not represent the full diversity of the global computing continuum market, although it is plausible for many scenarios.

\textbf{Construct Validity.} A potential threat to construct validity lies in the use of geographic distance as a proxy for latency when defining request constraints. While this is a common simplification in edge computing research \cite{mao2017survey}, it ignores other network factors such as congestion and routing hops that affect actual service-level objectives.

\textbf{Conclusion Validity.} To mitigate threats to conclusion validity, each scenario type was instantiated 100 times and results are reported as median execution times with their corresponding confidence intervals, reducing the impact of incidental topologies and transient system load. 

\section{Related Work}
\label{sec:relatedWork}

Resource allocation is arguably a densely studied topic; however, not if one approaches it from a new angle –by transferring well-established concepts from SaaS.

\subsection{Resource Allocation}

The question of \textit{where} and \textit{when} to allocate resources has long shaped modern cloud-based systems~\cite{vinothina2012survey}. It is typically formulated as a multi-objective optimization problem in which performance and operational criteria, such as latency, energy consumption, and monetary cost, must be jointly satisfied~\cite{Vergara2023}. These objectives are commonly formalized through a Service Level Agreement (SLA) between provider and consumer, requiring the provider to optimize resource allocation accordingly~\cite{sharma_sla_2023}. In computing continuum systems, however, this optimization gets more complex because it must also account for structural properties such as hierarchical deployment, geographically distributed resources, and node mobility~\cite{soumplis2022resource}. Consequently, numerous approaches have been proposed for continuum environments, e.g., by offloading tasks across the architecture~\cite{wu_intelligent_2023} or adapting device operation~\cite{sedlak_designing_2023}.

\subsection{Service Composition}

In parallel, service composition emerged as a core research area in service-oriented computing and web engineering. Early work~\cite{dustdar2005survey} emphasized the need for structured models to coordinate loosely coupled services, while more recent work~\cite{lemos2015web} formalized the key dimensions of composition. As service ecosystems grew, the focus shifted from purely functional composition to QoS-aware service selection~\cite{zeng2004qos}, where attributes such as latency, reliability, and cost constrain the search for an optimal binding. In distributed cloud environments, network-aware approaches further refined this model by distinguishing service-level QoS from network-level effects~\cite{wang2016towards}. Consequently, service composition—similar to resource allocation—has evolved into a multi-attribute optimization problem over alternative configurations.

\subsection{Toward a Unified Abstraction}

The computing continuum paradigm provides a unifying architectural perspective for these two research streams. By distributing computation across edge, fog, and cloud nodes \cite{moreschini2022cloud}, it spans multiple abstraction layers traditionally studied separately. Resource allocation is primarily addressed at the lower infrastructure layers (edge and fog), whereas service composition is typically investigated at the upper service layer of cloud-based software systems. 
However, from a structural perspective, both problems exhibit a similar formulation: selecting bindings within a constrained decision space governed by functional and non-functional attributes. This convergence suggests that the distinction between them is largely semantic rather than structural. Yet, despite this similarity, there is currently no shared modeling abstraction capable of representing variability across layers of the computing continuum.

This paper aims to merge these two areas by leveraging a structure traditionally used in the software-as-a-service domain –pricings \cite{garcia_tsc_2026}– to represent variability in resource allocation at the infrastructure level. By transposing this model across layers, we show that a representation originally conceived for service-level configuration can capture infrastructure-level decision spaces. This transfer enables the reuse of existing analysis tools, such as PRIME \cite{garcia_caise_2025}, and suggests that pricing models provide a promising approximation for expressing variability across the computing continuum.

\section{Conclusions \& Future Work}
\label{sec:conclusions}

In this paper, we investigated whether pricings can serve as a modeling abstraction for resource allocation in the computing continuum. In particular, we examined whether a pricing-based formulation is expressive enough to capture realistic multi-provider infrastructures with interoperability constraints and multi-mode nodes (\textbf{$RQ_1$}), whether such a formulation can be effectively employed to compute optimal bindings –namely, cost-optimal deployments under technical and economic constraints (\textbf{$RQ_2$}), and how the approach scales as demand intensity and topology size increase (\textbf{$RQ_3$}). 

Regarding \textbf{$RQ_1$}, the results provide strong evidence of the formulation’s expressiveness. All 96 deployment scenarios, spanning heterogeneous application types, varying infrastructure sizes, and distinct demand profiles, were successfully represented as iPricings without requiring ad-hoc adaptations. Concerning \textbf{$RQ_2$}, the proposed workflow consistently identified feasible and cost-optimal configurations whenever solutions existed, and correctly reported infeasibility otherwise. With respect to \textbf{$RQ_3$}, the empirical evaluation showed that execution time remains largely insensitive to demand growth and increases predictably with the number of candidate nodes. Nevertheless, for topologies involving up to two hundred nodes, solutions were computed in at most a few seconds.

Beyond these quantitative results, this work carries broader conceptual implications. Resource allocation and service composition have traditionally evolved as partially independent research streams, despite sharing a common structural foundation: both require exploring constrained decision spaces under multi-attribute objectives. By modeling infrastructure as a pricing, we blur the boundary between service-level and infrastructure-level variability. This suggests that pricings should not be regarded solely as commercial artifacts associated with SaaS offerings, but as \emph{general-purpose variability models} capable of structuring decision spaces across multiple layers of the computing continuum.

Several directions for future research emerge from this work. First, integrating runtime monitoring and dynamic re-optimization mechanisms would extend the current planning-oriented workflow toward adaptive resource management across the continuum. Second, validating the approach using real infrastructure specifications and production workloads would strengthen external validity. Third, from a modeling perspective, further generalization of the pricing abstraction could explore its application to other variability-intensive domains in distributed systems, such as federated learning orchestration or energy-aware scheduling. Finally, investigating hybrid solving strategies, decomposition techniques, or incremental optimization approaches may further improve scalability for infrastructures comprising thousands of candidate nodes.

\ifanonymized
\else
\section*{Acknowledgements}

This work has been partially supported by CNS2023-144359 and the European Union NextGenerationEU/PRTR under MICIU/AEI/10.13039/501100011033; other support comes from
grants 
PID2021-126227NB-C21, and 
PID2021-126227NB-C22,     
funded by MCIN / AEI / 10.13039 / 501100011033 / FEDER, EU ``ERDF a way of making Europe''; 
PID2024-155693NB-C44, 
funded by MICIU / AEI / 10.13039 / 501100011033 / FEDER, EU;
TED2021-131023B-C21 
funded by MCIN / AEI / 10.13039 / 501100011033, EU  “NextGenerationEU”/PRTR;
and the Horizonte 2027 doctoral training program funded by Junta de Andalucía (Consejería de Universidad, Investigación e Innovación).
\fi

\appendices
\section{Replicability \& Verifiability}

To support transparency, reproducibility, and benchmarking, we provide a research artifact in the form of a laboratory package that operationalizes the workflow presented in the paper. Beyond enabling the replication of our results, the artifact is intended as a reusable experimental asset for the research community, providing infrastructure topologies and demand scenarios that can serve as a common basis for future studies on resource allocation in the computing continuum.

\noindent The artifact, publicly available on Zenodo \cite{LABPACK}, includes:

\begin{enumerate}
    \item \textbf{Tool suite:} References to the repositories hosting the experimental tools and replication assets associated with the approach presented in the paper.

    \item \textbf{Source code snapshot:} Archived distributions containing the exact version of the Python package employed during experimentation. In contrast to the tool suite, which points to actively evolving repositories, these archived artifacts guarantee strict reproducibility by preserving the precise implementations used to obtain the reported results.

    \item \textbf{Replication scripts:} Automation utilities that simplify experiment configuration and execution, enabling the systematic reproduction of the analyses and evaluations described in the paper.

    \item \textbf{Technical documentation:} Detailed guidelines covering installation, configuration, and execution of the experimental workflow, as well as instructions to replicate the evaluation process end to end.

    \item \textbf{Experimental dataset:} A collection of 9{,}600 scenarios designed to assess the scalability of the approach.

    \item \textbf{Experiment configuration assets:} Parameterization files used to control scenario generation and experimental seeding, facilitating repeatability and enabling future benchmarking studies.
\end{enumerate}

The artifact is released exclusively for research purposes and should be regarded as a proof-of-concept implementation aimed at facilitating independent validation and further research on resource allocation in the computing continuum.

\bibliographystyle{IEEEtran}
\bibliography{references}

\end{document}